\documentclass[submission, Proceedings]{SciPost}

    \newcommand{\TUW}{2}
    \newcommand{\MPP}{3}
    \newcommand{\INFNRoma}{4}
    \newcommand{\INFNTorVergata}{5}
    \newcommand{\CNR}{6}
    \newcommand{\Sapienza}{7}
    \newcommand{\TUM}{8}
    \newcommand{\HEPHY}{9}
    \newcommand{\CEA}{1}
    \newcommand{\Ferrara}{10}
    \newcommand{\irfuAPC}{11}
    \newcommand{\LNGS}{12}
    \newcommand{\TorVergata}{13}
    \newcommand{\INFNFerrara}{14}
    \newcommand{\CIUC}{15}

\hypersetup{
    colorlinks,
    linkcolor={red!50!black},
    citecolor={blue!50!black},
    urlcolor={blue!80!black}
}

\usepackage[bitstream-charter]{mathdesign}
\DeclareSymbolFont{usualmathcal}{OMS}{csy}{m}{n}
\DeclareSymbolFontAlphabet{\mathcal}{usualmathcal}

\begin{document}

\begin{center}{\Large \textbf{
Exploring coherent elastic neutrino-nucleus scattering of reactor neutrinos with the NUCLEUS experiment\\
}}\end{center}

\begin{center}
C.~Goupy\textsuperscript{\CEA, $\star$}, 
H.~Abele\textsuperscript{\TUW},
G.~Angloher\textsuperscript{\MPP}, 
A.~Bento\textsuperscript{\MPP,\CIUC}, 
L.~Canonica\textsuperscript{\MPP}, 
F.~Cappella\textsuperscript{\INFNRoma}, 
L.~Cardani\textsuperscript{\INFNRoma}, 
N.~Casali\textsuperscript{\INFNRoma}, 
R.~Cerulli\textsuperscript{\INFNTorVergata,\TorVergata}, 
I.~Colantoni\textsuperscript{\CNR,\INFNRoma}, 
A.~Cruciani\textsuperscript{\INFNRoma}, 
G.~Del~Castello\textsuperscript{\Sapienza,\INFNRoma}, 
M.~del~Gallo~Roccagiovine\textsuperscript{\Sapienza,\INFNRoma}, 
A.~Doblhammer\textsuperscript{\TUW}, 
S.~Dorer\textsuperscript{\TUW}, 
A.~Erhart\textsuperscript{\TUM}, 
M.~Friendl\textsuperscript{\HEPHY}, 
A.~Garai\textsuperscript{\MPP}, 
V.M.~Ghete\textsuperscript{\HEPHY}, 
D.~Hauff\textsuperscript{\MPP}, 
F.~Jeanneau\textsuperscript{\CEA}, 
E.~Jericha\textsuperscript{\TUW}, 
M.~Kaznacheeva\textsuperscript{\TUM}, 
A.~Kinast\textsuperscript{\TUM}, 
H.~Kluck\textsuperscript{\HEPHY}, 
A.~Langenk\"{a}mper\textsuperscript{\TUM}, 
T.~Lasserre\textsuperscript{\irfuAPC}, 
D.~Lhuillier\textsuperscript{\CEA}, 
M.~Mancuso\textsuperscript{\MPP}, 
B.~Mauri\textsuperscript{\CEA}, 
A.~Mazzolari\textsuperscript{\INFNFerrara}, 
E.~Mazzucato\textsuperscript{\CEA}, 
H.~Neyrial\textsuperscript{\CEA}, 
C.~Nones\textsuperscript{\CEA}, 
L.~Oberauer\textsuperscript{\TUM}, 
T.~Ortmann\textsuperscript{\TUM}, 
L.~Pattavina\textsuperscript{\LNGS,\TUM}, 
L.~Peters\textsuperscript{\TUM}, 
F.~Petricca\textsuperscript{\MPP}, 
W.~Potzel\textsuperscript{\TUM}, 
F.~Pr\"{o}bst\textsuperscript{\MPP}, 
F.~Pucci\textsuperscript{\MPP}, 
F.~Riendl\textsuperscript{\HEPHY,\TUW}, 
R.~Rogly\textsuperscript{\CEA}, 
M.~Romagnoni\textsuperscript{\INFNFerrara}, 
J.~Rothe\textsuperscript{\TUM}, 
N.~Schermer\textsuperscript{\TUM}, 
J.~Schieck\textsuperscript{\HEPHY,\TUW}, 
S.~Sch\"{o}nert\textsuperscript{\TUM}, 
C.~Schwertner\textsuperscript{\HEPHY,\TUW}, 
L.~Scola\textsuperscript{\CEA},
G.~Soum-Sidikov\textsuperscript{\CEA}, 
L.~Stodolsky\textsuperscript{\MPP}, 
R.~Strauss\textsuperscript{\TUM}, 
M.~Tamisari\textsuperscript{\Ferrara, \INFNFerrara}, 
C.~Tomei\textsuperscript{\INFNRoma}, 
M.~Vignati\textsuperscript{\Sapienza, \INFNRoma}, 
M.~Vivier\textsuperscript{\CEA}, 
V.~Wagner\textsuperscript{\TUM}, 
A.~Wex\textsuperscript{\TUM}
\end{center}

\begin{center}
{\bf \CEA} IRFU, CEA, Universit\'{e} Paris Saclay, F-91191 Gif-sur-Yvette, France 
\\
{\bf \TUW} Atominstitut, Technische Universit\"at Wien, A-1020 Wien, Austria
\\
{\bf \MPP} Max-Planck-Institut f\"ur Physik, D-80805 M\"unchen, Germany
\\
{\bf \INFNRoma} Istituto Nazionale di Fisica Nucleare -- Sezione di Roma, I-00185 Roma, Italy
\\
{\bf \INFNTorVergata} Istituto Nazionale di Fisica Nucleare -- Sezione di Roma "Tor Vergata", I-00133 Roma, Italy
\\
{\bf \CNR} Consiglio Nazionale delle Ricerche, Istituto di Nanotecnologia, I-00185 Roma, Italy
\\
{\bf \Sapienza} Dipartimento di Fisica, Sapienza Universit\`{a} di Roma, I-00185 Roma, Italy
\\
{\bf \TUM} Physik-Department, Technische Universit\"at M\"unchen, D-85748 Garching, Germany
\\
{\bf \HEPHY} Institut f\"ur Hochenergiephysik der \"Osterreichischen Akademie der Wissenschaften, A-1050 Wien, Austria
\\
{\bf \Ferrara} Dipartimento di Fisica, Universit\`{a} di Ferrara, I-44122 Ferrara, Italy
\\
{\bf \irfuAPC} IRFU (DPhP \& APC), Universit\'{e} Paris-Saclay, F-91191 Gif-sur-Yvette, France
\\
{\bf \LNGS} Istituto Nazionale di Fisica Nucleare -- Laboratori Nazionali del Gran Sasso, I-67100 Assergi (L’Aquila), Italy
\\
{\bf \TorVergata} Dipartimento di Fisica, Universit\`{a} di Roma "Tor Vergata", I-00133 Roma, Italy
\\
{\bf \INFNFerrara} Istituto Nazionale di Fisica Nucleare -- Sezione di Ferrara, I-44122 Ferrara, Italy
\\
{\bf \CIUC} CIUC, Departamento de Fisica, Universidade de Coimbra, P3004 516 Coimbra, Portugal
\\
\vspace{0.3cm}
$\star$ chloe.goupy@cea.fr
\end{center}

\begin{center}
\today
\end{center}

\definecolor{palegray}{gray}{0.95}
\begin{center}
\colorbox{palegray}{
  \begin{tabular}{rr}
  \begin{minipage}{0.1\textwidth}
    \includegraphics[width=30mm]{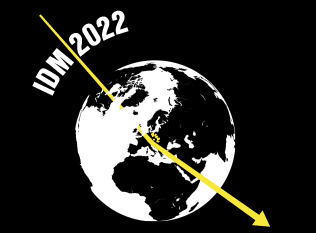}
  \end{minipage}
  &
  \begin{minipage}{0.85\textwidth}
    \begin{center}
    {\it 14th International Conference on Identification of Dark Matter}\\
    {\it Vienna, Austria, 18-22 July 2022} \\
    \doi{10.21468/SciPostPhysProc.?}\\
    \end{center}
  \end{minipage}
\end{tabular}
}
\end{center}

\section*{Abstract}
{\bf
The NUCLEUS experiment aims to perform a high-precision measurement of Coherent Elastic Neutrino–Nucleus Scattering (CEvNS) at the EdF Chooz B nuclear power plant in France. CEvNS is a unique process to study neutrino properties and to search for physics beyond the Standard Model. The study of CEvNS is also important for light Dark-Matter searches. It could be a possible irreducible background for high-sensitivity Dark-Matter experiments. NUCLEUS is an experiment under construction based on ultra-low threshold ($\sim$20 eV$_{nr}$) cryogenic calorimeters, operated at tens-of-mK temperatures. 

}


\section{Introduction}
\label{sec:Intro}
The NUCLEUS experiment aims to detect Coherent Elastic Neutrino-Nucleus Scattering (CEvNS) from reactor anti-neutrinos. Created by $\beta$-decays of the fission products, they have energies up to 10MeV such that they interact by neutral current with the nucleus as a whole. This is the so-called "fully coherent regime" of CEvNS. 
Moreover nuclear reactors, with $\mathcal{O}(10^{20})$ neutrinos produced per GW$_{th}$ and per second, are copious sources of neutrinos, making them promising sources to observe CEvNS. They offer a way which is complementary to experiments running at stopped-pion sources (e.g. COHERENT \cite{Coherent}) which study CEvNS in a higher energy regime.
\par CEvNS is a new probe to test the Standard Model and to search for new physics. It can be used to look for the sterile neutrino, study electromagnetic properties of the neutrino \cite{EM_Vogel}, measure the Weinberg angle at low momentum transfer \cite{CEvNS_Drukier} or look for non standard interactions \cite{Scholberg}. The measurement of CEvNS is closely connected to Dark Matter searches. Both are sharing the same experimental signature: a sub-keV nuclear recoil and a similar detection technique based on cryogenic bolometers. The synergy between DM searches and CEvNS expands to the characterisation of "the neutrino floor", which will ultimately limit the sensitivity of the current and next generation of experiments in the low mass regime.

\par CEvNS is a flavor independent and thresholdless process. 
The cross-section may exhibit up to two orders of magnitude boost, in comparison with the Inverse Beta Decay cross-section, from its dependency upon the squared number of neutrons in the target nucleus. As a consequence, the heavier the target, the higher the cross-section, but the lower the recoil energy.
Therefore, the choice of the target material should find a good compromise between high counting rate and detectable nuclear recoils. 
The challenges to observe CEvNS from reactor neutrinos are similar to the ones tackled by rare event Dark Matter search experiments in the low WIMP mass regime. First, the need to reach a low nuclear recoil threshold and second, a good background mitigation. The shallow overburden often imposed by the location of nuclear reactors is an additional challenge for reactor neutrino experiments and requires a good shielding strategy to discriminate cosmic-rays induced backgrounds that become dominant above ground.
\par The NUCLEUS experiment will use gram-scaled cryogenic calorimeters to detect CEvNS from reactor neutrinos at the Chooz B Nuclear Power Plant in France. The detection principle and the experimental apparatus are described in the following section.

\section{The NUCLEUS experiment}
\label{sec:NUCLEUS}
\par The NUCLEUS experimental site, called the "Very Near Site", is a 25~m$^2$ room in the basement of an administrative building located at 102~m and 72~m from the two 4.25~GW$_\text{th}$ reactor cores of the Chooz B nuclear Power Plant in France. This location guarantees a high neutrino average flux of about $1.7\times10^{12}$~$\nu$/(s cm$^2$). Nevertheless, its low overburden (3 meters water equivalent) implies the need of a relevant cosmic-ray induced background mitigation strategy \cite{nucleus2019}. Several layers of passive and active shielding are necessary in order to reach a background level of 100~counts/(kg~day~keV). In the region of interest, between 20~eV and 100~eV, a total counting rate of 30~$\nu$/(kg day) above background is expected in the NUCLEUS experiment.

\begin{figure}[h]
\centering
\includegraphics[width=1\textwidth]{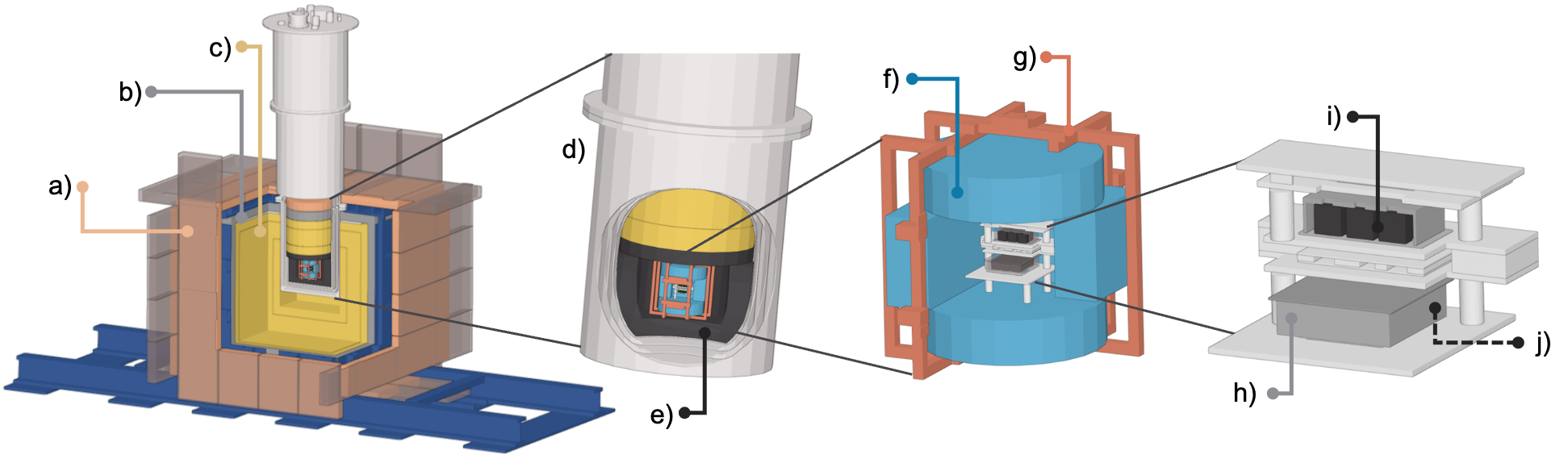}
\caption{CAD drawing of the NUCLEUS experiment. \textbf{From left to right}, the full NUCLEUS apparatus with the shielding mechanical structure (dark blue), \textbf{a)} 28 5-cm thick Muon Veto panels, \textbf{b)} a 5-cm thick lead layer, and \textbf{c)} a 20-cm thick borated polyethylene. \textbf{d)} A dilution refrigerator is inserted inside the shielding and contains \textbf{e)} a 4-cm thick boron carbide layer and \textbf{f)} a Cryogenic Outer Veto made of six high purity germanium crystals held by \textbf{g)} a copper cage. Finally the cryogenic detectors are organised in two arrays of nine cubes of \textbf{i)} CaWO$_4$ and \textbf{j)} Al$_2$O$_3$, held by \textbf{h)} the silicon inner veto.}
\label{figure1}
\end{figure}

\subsection{The NUCLEUS target detectors: gram-scaled cryogenic calorimeters}
Gram-scaled cryogenic calorimeters made of Sapphire (Al$_2$O$_3$) and Calcium Tungstate (CaWO$_4$), will be used, totaling 10~g of detectors. They are based on the technology developed by CRESST for Dark Matter searches \cite{cresst}. When a particle induces a nuclear recoil in the target crystal it creates a lattice vibration which is travelling along the crystal through phonons. This lattice vibration is read out by a Transition Edge Sensor made of Tungsten (W-TES). The detector is operated at the transition temperature of the Tungsten film, so the slight rise of temperature induced by phonons means a significant rise of the TES resistance which can be read-out by a SQUID (Superconducting QUantum Interference Device) \cite{Rothe:2021}. A prototype made of 0.5~g of Al$_2$O$_3$ showed the first proof of principle by reaching a threshold of E$_\text{th} = (19.7 \pm  0.9)$ ~eV$_\text{nr}$ \cite{nucleus_proto}.
\par The NUCLEUS experiment is taking advantage of the N$^2$-dependency of CEvNS cross section by using a multi-target approach (see figure \ref{figure2}a) with an array of 9 CaWO$_4$ cubes (6~g) to measure CEvNS and an array of 9 Al$_2$O$_3$ (4~g) to perform an in-situ background measurement \cite{nucleus2019}. In July 2022, 18 CaWO$_4$ individual cubes have already been equipped with W-TES and have been successfully tested. Comparable performances have been achieved in the NUCLEUS dry refrigerator with respect to the early prototype operated in a wet fridge.

\subsection{The NUCLEUS active and passive shielding layers}
\par
\begin{figure}[h]
\centering
\includegraphics[width=1\textwidth]{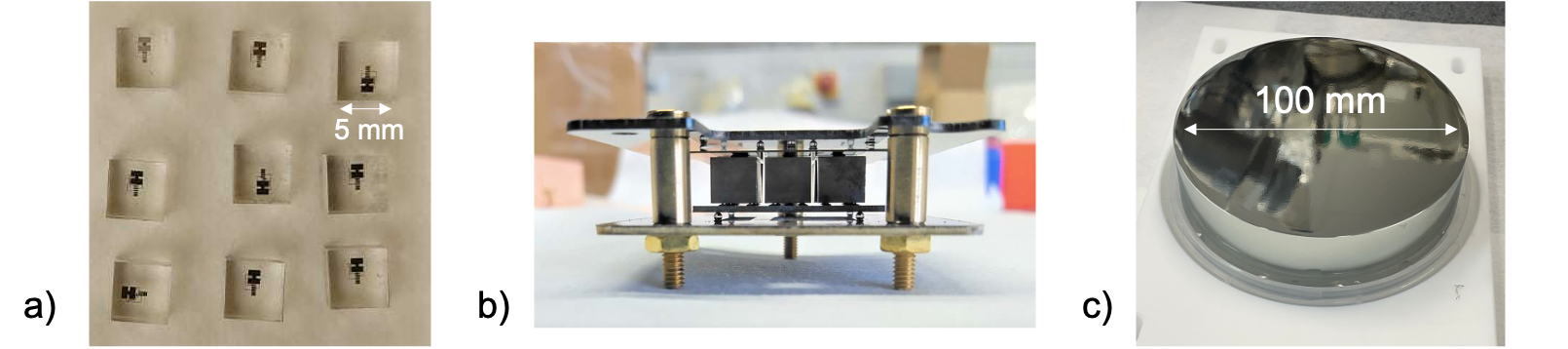}
\caption{Photographs of NUCLEUS detector crystals and IV prototype. \textbf{a)} Nine CaWO$_4$ cubes cut and equipped with W-TES. \textbf{b)} Inner Veto mock-up with nine Si detector dummies. \textbf{c)} High Purity Ge crystal of the future NUCLEUS Cryogenic Outer Veto.}
\label{figure2}
\end{figure}

 The crystals will be held by an TES-instrumented silicon structure called Inner Veto (IV). Operated in anti-coincidence with the cryogenic detectors, it will reject surface and mechanical-stress relaxation related events. A first mock-up (see figure \ref{figure2}b) has been mechanically and thermally tested with Si detector dummies. New tests will soon be performed with the CaWO$_4$ crystals.
\par The next veto system is called Cryogenic Outer Veto (COV) and is made of six 2.5-cm thick high purity germanium crystals read out in ionisation mode with a threshold of $\mathcal{O}(10~\text{keV})$. They form a 4$\pi$-covering active shielding against external backgrounds (mainly against ambient radioactivity and atmospheric muons). Tests with a prototype proved the possibility to use it in anti-coincidence mode with a bolometric detector \cite{Beatrice}. The two final cylindrical crystals have already been prepared, tested and validated. The four rectangular ones are under production and will soon be tested as well. The COV and the cryogenic detectors will be held by a copper cage suspended to a spring system to decouple the detectors from the cryostat vibrations \cite{Nicole}. \newline

\par The most external active shielding system is located at room temperature. It is made of 28 5-cm thick scintillating plastics read out with optical fibers and Silicon PhotoMultipliers (SiPMs) and acts as a muon veto. A module prototype of the Muon Veto (MV) has been completely validated and characterised \cite{MV}. The final modules are under production and will soon be integrated to the structure. In order to reach a geometrical efficiency $>99\%$, the coverage of the MV is increased to 4$\pi$ by adding a Cryogenic Muon Veto at the 800mK stage inside the cryostat. First measurements proved its efficiency without loss of scintillation in cold conditions \cite{Andi}.
\par In addition to the three active vetos, passive shields are covering the detector in both cryogenic and external stages. First, a 5-cm thick lead layer is used to shield the gammas from ambient radioactivity. A 20-cm thick 5\% borated polyethylene layer moderates and attenuates neutrons. Finally, an up-to-4-cm thick boron carbide (B$_4$C) layer will be added inside the cryostat, the closest to the detectors. Its role is to capture slow and thermal neutrons reaching the vicinity of the detectors provoked by the interaction of fast atmospheric neutrons with the several shielding layers of the experiment. The importance of such a layer has been demonstrated by Monte Carlo simulations \cite{SimuPoster}.

\section{First estimate of NUCLEUS background}
In order to estimate the background level in the target crystals, extensive Geant 4 \cite{G4} Monte Carlo simulations are run. The full NUCLEUS apparatus geometry has been implemented as well as the vetoes anti-coincidence cuts with their respective thresholds (30~eV for the IV, 1~keV for the COV, 5~MeV for the MV). In addition, an anti-coincidence between the crystals is applied with a 10~eV threshold. Atmospheric muons, atmospheric neutrons and ambient gammas have been identified as the main contributors of background for NUCLEUS. A first set of simulations permitted to optimise the shielding geometry. Preliminary results demonstrated a good mitigation of the expected background from the NUCLEUS shielding geometry and the possibility to reach the goal level of <100~counts/(kg~day~keV) \cite{SimuPoster}.
\par Many rare event search experiments observe an exponentially rising background at sub-keV energies close to their respective threshold. This rise still has an unknown origin and currently overwhelms our knowledge about external backgrounds in this energy regime \cite{Excess}. Even if this excess represents a challenge for the NUCLEUS targeted background level, the previously described background mitigation strategy will allow investigating its origin and nature.

\section{Conclusion and outlook}
The NUCLEUS experiment aims at observing CEvNS from reactor neutrinos at the Chooz B Nuclear Power Plant using gram-scaled cryogenic calorimeters with a 20~eV$_{nr}$ threshold. MC simulations showed the ability of the nearly 4$\pi$-covering active and passive shielding layers to mitigate the cosmic and radiogenic background components to meet the goal of 100~counts/(kg~day~keV) in the ROI.
\par The blank assembly phase is now proceeding at the underground laboratory of the Technical University of Munich, with a commissioning phase likely to start at the beginning of next year. The goal of this phase is first, to test the mechanical integration but also to perform calibrations at keV energies and below with LED systems, XRF and neutron source (with the CRAB project \cite{CRAB, Vici}). The installation at the Chooz complex is then expected to happen at the end of 2023 to be ready to start the physics run beginning of 2024. This first phase of NUCLEUS with 10~g of detectors aims to perform a first measurement of CEvNS. A second phase with a larger detector (1~kg) is expected to come later in order to achieve a precision measurement of CEvNS cross-section at the several percent level.

\section*{Acknowledgements}
This work has been financed by the CEA, the INFN, the ÖAW and partially supported by the TU Munich and the MPI für Physik. NUCLEUS members acknowledge additional funding by the DFG through the SFB1258 and the Excellence Cluster ORIGINS, by the European Commission through the ERC-StG2018-804228 “NU-CLEUS”, by the P2IO LabEx (ANR-10-LABX-0038) in the framework "Investissements d’Avenir" (ANR-11-IDEX-0003-01) managed by the Agence Nationale de la Recherche (ANR), France and by the Austrian Science Fund (FWF) through the "P 34778-N, ELOISE‘“.





\bibliography{bibliography.bib}

\nolinenumbers

\end{document}